\documentclass[aps,pre,twocolumn,showpacs,longbibliography,superscriptaddress,final,floatfix]{revtex4-2}
\usepackage{amsfonts}
\usepackage{amsmath}
\usepackage{amssymb}
\usepackage{latexsym}
\usepackage{xcolor}
\usepackage{bm}
\usepackage{relsize}
\usepackage{float}
\usepackage{subfloat}
\usepackage{dcolumn}
\usepackage{epsfig}
\usepackage{graphicx}
\usepackage[breaklinks,colorlinks = true,linkcolor = red,urlcolor=blue,citecolor=red]{hyperref}
\usepackage{multirow}
\usepackage{makecell}
\usepackage[bbgreekl]{mathbbol}

\begin{document}

\title{Quantum Langevin dynamics of a charged particle in a magnetic field : Response function, position-velocity and velocity autocorrelation functions}
\author{Suraka Bhattacharjee}
 \affiliation{Raman Research Institute, Bangalore-560080, India} 
\author{Urbashi Satpathi}
 \affiliation{International Center for Theoretical Sciences, Tata Institute of Fundamental Research, Bangalore-560089, India}
 \author{Supurna Sinha}
 \affiliation{Raman Research Institute, Bangalore-560080, India}

\date{\today}

\begin{abstract}
We use the Quantum Langevin equation as a starting point to study the response function, the position-velocity correlation function and the velocity autocorrelation function of a charged Quantum Brownian particle in the presence of a magnetic field and linearly coupled to a heat bath via position coordinate. We study two bath models- the Ohmic bath model and the Drude bath model and make a detailed comparison in various time-temperature regimes. 
For both bath models there is a competition between the cyclotron frequency and the viscous damping rate giving rise to a transition from an
oscillatory to a monotonic behaviour as the damping rate is increased. In the zero point fluctuation dominated low temperature regime, non-trivial noise correlations lead to some interesting features in this transition. We study the role of the memory time scale which comes into play in 
the Drude model and study the effect
of this additional time scale. We discuss the experimental 
implications of our analysis in the context of experiments in cold ions.

\end{abstract}

\maketitle

\section{Introduction}

The response function of a system, which 
measures the response of a system to an 
external perturbation characterises the intrinsic properties of the system for instance, electric polarisability, magnetic susceptibility 
and so on\cite{balescu}. 
Thus, the response function of a system 
is of central importance in the realm of 
non-equilibrium statistical mechanics\cite{mazenko}.

In recent years there have been theoretical and experimental research on diffusion in oscillatory and damped regimes. In particular, Bloch oscillations and their damping due to spontaneous emission have been studied in the context of cold atoms in optical lattices \cite{Kolovsky,Dahan}. 
 
 In contrast to earlier work discussed above where the authors address the issue of Brownian motion of neutral particles, here we consider a charged particle in the presence of a magnetic field in
a viscous environment. There has been a study of the classical Langevin dynamics of a charged particle in a magnetic field in the high temperature classical domain\cite{Paraan}. 
In contrast, in this paper our focus is on an analysis of   
the position response function of a charged 
particle in a viscous environment in the 
presence of a magnetic field via the Quantum 
Langevin Equation. 
The system under consideration is characterised
by two competing rates:
the cyclotron frequency $\omega_c= qB/{mc}$, where 
$m$ is the mass of the particle, $q$ is the 
charge, $c$ is the speed of light, $B$ is the 
strength of the magnetic field and $\gamma$, the rate associated with dissipation. 
In Ref. \cite{sushantapramana} the authors do 
analyse such a system. However, we go beyond
that study and explore
various time regimes and analyse in detail 
the onset of 
oscillatory response of this system for two different bath models - the Ohmic bath and the 
Drude bath models. In the Drude bath model there is an 
additional time scale, the memory time $\tau$ which 
introduces some interesting quantitative effects. We also study the behaviour of the position-velocity correlation function and the velocity autocorrelation function in various time-temperature regimes for the two bath models. \\
The paper is organised as follows. In Sec $II$ we introduce the Quantum Langevin Equation and define the response function, the position-velocity correlation function and the velocity autocorrelation function. In Sec $III $ we analyse the position response function
for the Ohmic model and the Drude model. In Sec $IV$ we analyse the position-velocity correlation function for the two models and in Sec $V$ we analyse the velocity autocorrelation function for the system for the two bath models. In Sec $VI$ we discuss the results of our study and compare the behaviour of the response function, the position-velocity correlation function and the velocity autocorrelation function for the two bath models at various time temperature regimes. In Sec $VII$ we discuss the experimental implications of our study and finally end the paper in Sec $VIII$ with some concluding remarks.

\section{Position correlation function in the presence of a magnetic field}

The Hamiltonian for the motion of a charged quantum Brownian particle in the presence of a magnetic field and linearly coupled via position coordinate to a passive heat bath, characterized by the bath in thermal equilibrium is given by \cite{Li1,Li2}:
\begin{align}
\notag H=\frac{1}{2m}\left(p-\frac{qA}{c}\right)^2+ \sum_j \left[\frac{p_j^2}{2m_j}+\frac{1}{2}m_j\omega_j^2 \left(q_j-r \right)^2 \right]\\    \label{Hamiltonian}
\end{align} 
where $m, q, r, p$ are the mass, charge, position and momentum of the particle respectively, whereas, $m_j, q_j, p_j$ and $\omega_j$ are the mass, position co-ordinate, momentum co-ordinate and frequency of the $j^{th}$ oscillator in the bath respectively. $A(r)$ represents the vector potential corresponding to the applied magnetic field. \\
We derive the Quantum Langevin equation (QLE) for this system.
We outline below the basic steps used in the derivation of the QLE:\\
(i) We first obtain the Heisenberg equations of motion for the heat bath and the system of the charged particle linearly coupled to the heat bath via position coordinate. 
 We then solve these equations for the bath variables, and substitute the solution into the equations for the charged particle to obtain a reduced description of the particle motion.
  The solution contains explicit expressions for the dynamical variables at time t in terms of their initial values.
(ii) We make specific assumptions about the initial
state of the system. For instance we  assume that the heat bath was at thermal equilibrium, which is satisfied by the fourier transform of the memory function being a positive real function \cite{Li1}.\\
Thus, the generalized quantum Langevin equation is formulated from the Hamiltonian (Eq.\ref{Hamiltonian}), where, the effect of the passive heat bath is only retained in the memory kernel and the random fluctuating force \cite{Li1,Weiss,Ford1,satpathi2018quantum}:
\begin{eqnarray}
\notag m \ddot{\vec{r}}(t)=-\int\mu (t-t')\dot{\vec{r}}(t')dt'+\frac{q}{c}(\dot{\vec{r}}(t)\times\vec{B})+\vec{F}(t) \\
\end{eqnarray}
where, $ m $ is the mass of the particle, $ \mu(t) $ is the memory kernel, $ q $ is the charge, $ c $ is the speed of light, $ \vec{B} $ is the applied magnetic field and $ \vec{F}(t) $ is the random force with
the following properties :
\begin{align}
\langle{F_{\alpha}(t)}\rangle=0
\end{align}

\begin{align}
\notag \frac{1}{2}\langle{\lbrace F_{\alpha}(t),F_{\beta}(0)}\rbrace\rangle =\frac{\delta_{\alpha \beta}}{2 \pi}\int_{-\infty}^\infty{{d\omega}Re[\mu(\omega)]}
\hbar\omega  \\ \hspace{0.2cm} \smaller{\times}  \coth(\frac{\hbar\omega}{2k_BT}) e^{-i\omega t} \label{noisecor}
\end{align}

\begin{equation}
\langle{[F_{\alpha}(t),F_{\beta}(0)}]\rangle=\frac{\delta_{\alpha \beta}}{ \pi}\int_{-\infty}^\infty{{d\omega}Re[\mu(\omega)]}
\hbar\omega e^{-i\omega t}\\
\end{equation}
Here $\alpha, \beta = x, y, z$ and $\delta_{\alpha \beta}$ is the Kronecker delta function.
Also 
$\mu(\omega)= \int_{-\infty}^{\infty}{dt \mu(t) e^{i\omega t}}$.

We consider an uniform magnetic field along the $z$ axis. This results in the following solutions to the motion of the charged particle in the $x-y$ plane:  \cite{satpathi2018quantum} 
\begin{align}
\tilde{x}(\omega)=\frac{1}{m} \frac{i\omega_c \tilde{F}_y (\omega)-(\omega-iK(\omega))\tilde{F}_x(\omega)}{\omega[\omega^2-\omega_c^2-\tilde{K}(\omega)^2-2i\omega \tilde{K}(\omega)]} \\
\tilde{y}(\omega)=\frac{1}{m} \frac{-i\omega_c \tilde{F}_x (\omega)-(\omega-iK(\omega)){F}_y(\omega)}{\omega[\omega^2-\omega_c^2-\tilde{K}(\omega)^2-2i\omega \tilde{K}(\omega)]} 
\end{align}
Here, $ \omega_{c}=\frac{eB}{mc}$ is the cyclotron frequency, $ K(\omega)=\frac{\mu(\omega)}{m}$ . 
Using properties of the random force \cite{ford}, we can write the position autocorrelation function for the $x,y$ components \cite{satpathi2018quantum}, 
 \begin{widetext}
  \begin{eqnarray}
 C_{x}(t)&=&\frac{1}{2}\langle \left\lbrace x(t),x(0) \right\rbrace \rangle\nonumber\\
&=&\frac{\hbar}{2\pi m}\int_{-\infty}^{\infty}d\omega \mathrm{Re}[K(\omega)] \frac{\left[\left(\omega + \mathrm{Im}[K(\omega)]\right)^{2}+\omega_{c}^{2}+\mathrm{Re}[K(\omega)]^{2} \right]\mathrm{coth}\left(\frac{\hbar\omega}{2k_{B}T} \right)e^{-i\omega t}}{\omega\left\lbrace  \left[\left(\omega + \mathrm{Im}[K(\omega)]\right)^{2}+\omega_{c}^{2}+\mathrm{Re}[K(\omega)]^{2}  \right]^{2}- 4\omega_c^{2}\left(\omega + \mathrm{Im}[K(\omega)]\right)^{2} \right\rbrace } \label{positioncorrelation}
 \end{eqnarray}
 \end{widetext}
The same expression is obtained for $C_{y}(t)=\frac{1}{2}\langle \left\lbrace y(t),y(0) \right\rbrace\rangle$. Using the position autocorrelation function one can get various physical observables like the mean square displacement, response function, position-velocity correlation function and velocity autocorrelation function. In Ref. \cite{satpathi2018quantum}, the mean square displacement has been analysed in detail. In the following subsections we will discuss the expressions for the response function, position-velocity correlation function and velocity autocorrelation function, all evaluated using the position correlation function (defined in Eq. (\ref{positioncorrelation})) as a starting point. We analyse the behaviour of these observables in various time temperature regimes for two bath models- the Ohmic model and the Drude model. 
\vspace{2cm}
\subsection{Response function}
The response function $R(t)$ pertaining to an external perturbation f(t) is given by:
\begin{equation}
    <x(t)>=\int R(t-t')f(t')dt'
\end{equation}
where $<x(t)>$ is the expectation value of the position
of the particle. 
In the Fourier domain the response function R$_x(\omega$) can be expressed in terms of the position correlation function C$_x(\omega)$ as follows:
\cite{balescu,Peter}
\begin{eqnarray}
\mathrm{Im} R_{x}(\omega)&=&\frac{1}{\hbar}\tanh \left(\frac{\hbar \omega}{2k_B T}\right)C_{x}(\omega)\label{FDT}
\end{eqnarray}
From Eq. (\ref{positioncorrelation}), we get
\begin{widetext}
\begin{eqnarray}
C_{x}(\omega)&=&\frac{\hbar}{ m} \mathrm{Re}[K(\omega)] \frac{\left[\left(\omega + \mathrm{Im}[K(\omega)]\right)^{2}+\omega_{c}^{2}+\mathrm{Re}[K(\omega)]^{2} \right]\mathrm{coth}\left(\frac{\hbar\omega}{2k_{B}T} \right)}{\omega\left\lbrace  \left[\left(\omega + \mathrm{Im}[K(\omega)]\right)^{2}+\omega_{c}^{2}+\mathrm{Re}[K(\omega)]^{2}  \right]^{2}- 4\omega_c^{2}\left(\omega + \mathrm{Im}[K(\omega)]\right)^{2} \right\rbrace } \label{positioncorrelstionomega}
\end{eqnarray}
Using Eqs. (\ref{FDT}) and (\ref{positioncorrelstionomega}), we get
\begin{eqnarray}
\mathrm{Im}R_{x}(\omega)&=&\frac{1}{ m} \mathrm{Re}[K(\omega)] \frac{\left[\left(\omega + \mathrm{Im}[K(\omega)]\right)^{2}+\omega_{c}^{2}+\mathrm{Re}[K(\omega)]^{2} \right]}{\omega\left\lbrace  \left[\left(\omega + \mathrm{Im}[K(\omega)]\right)^{2}+\omega_{c}^{2}+\mathrm{Re}[K(\omega)]^{2}  \right]^{2}- 4\omega_c^{2}\left(\omega + \mathrm{Im}[K(\omega)]\right)^{2} \right\rbrace } \label{responsefunctionomega}
\end{eqnarray}
\end{widetext}
Knowing $\mathrm{Im}R_{x}(\omega)$, one can get the expression for $\mathrm{Re}R_{x}(\omega)$ using the Kramers Kronig relation\cite{bohren2010did}
\begin{eqnarray}
\mathrm{Re}R_{x}(\omega)&=&\frac{1}{\pi}P\int_{-\infty}^{\infty}\frac{\omega' \mathrm{Im}R_{x}(\omega')}{(\omega'^2-\omega^2)}d\omega'\label{kramers}
\end{eqnarray}
where `P' refers to the Principal value of the integral. \\
Collecting the real and imaginary parts we get, $R_x (\omega)=\mathrm{Re}R_{x}(\omega)+i\mathrm{Im}R_{x}(\omega)$, the Fourier transform of which gives the time dependent position response function $R_x(t)$,
\begin{eqnarray}
R_x (t)&=&\frac{1}{2\pi}\int_{-\infty}^{\infty}R_x (\omega) e^{-i\omega t}=\mathcal{F}\left[R_x (\omega)\right]\\
&=&\mathcal{F}\left[\mathrm{Re}R_{x}(\omega)\right]+i\mathcal{F}\left[\mathrm{Im}R_{x}(\omega)\right]\label{responset}
\end{eqnarray}
\subsection{Position-velocity correlation function}
The position-velocity correlation function is defined as
\begin{eqnarray}
C_{xv_{x}}(t)&=&\frac{1}{2}\langle \left\lbrace x(t),v_x(0) \right\rbrace\rangle=\frac{1}{2}\langle \left\lbrace x(0),v_x(-t) \right\rbrace\rangle\nonumber\\
&=&\frac{d}{dt}\frac{1}{2}\langle \left\lbrace x(t),x(0) \right\rbrace\rangle=\frac{d}{dt}C_{x}(t)
\end{eqnarray}
Using Eq. (\ref{positioncorrelation}), we get,
 \begin{widetext}
  \begin{eqnarray}
 C_{xv_{x}}(t)&=&\frac{-i\hbar}{2\pi m}\int_{-\infty}^{\infty}d\omega \mathrm{Re}[K(\omega)] \frac{\left[\left(\omega + \mathrm{Im}[K(\omega)]\right)^{2}+\omega_{c}^{2}+\mathrm{Re}[K(\omega)]^{2} \right]\mathrm{coth}\left(\frac{\hbar\omega}{2k_{B}T} \right)e^{-i\omega t}}{\left\lbrace  \left[\left(\omega + \mathrm{Im}[K(\omega)]\right)^{2}+\omega_{c}^{2}+\mathrm{Re}[K(\omega)]^{2}  \right]^{2}- 4\omega_c^{2}\left(\omega + \mathrm{Im}[K(\omega)]\right)^{2} \right\rbrace } \label{positionvelocitycorrelation}
 \end{eqnarray}
 \end{widetext}
 
\subsection{Velocity autocorrelation function}
The velocity autocorrelation function is defined as
\begin{eqnarray}
C_{v_x}(t)&=&\frac{1}{2}\langle \left\lbrace v_x(t),v_x(0) \right\rbrace\rangle=-\frac{d^2}{dt^{2}}C_{x}(t)
\end{eqnarray}
 \begin{widetext}
  \begin{eqnarray}
 C_{v_{x}}(t)&=&\frac{\hbar}{2\pi m}\int_{-\infty}^{\infty}d\omega \;\omega \mathrm{Re}[K(\omega)] \frac{\left[\left(\omega + \mathrm{Im}[K(\omega)]\right)^{2}+\omega_{c}^{2}+\mathrm{Re}[K(\omega)]^{2} \right]\mathrm{coth}\left(\frac{\hbar\omega}{2k_{B}T} \right)e^{-i\omega t}}{\left\lbrace  \left[\left(\omega + \mathrm{Im}[K(\omega)]\right)^{2}+\omega_{c}^{2}+\mathrm{Re}[K(\omega)]^{2}  \right]^{2}- 4\omega_c^{2}\left(\omega + \mathrm{Im}[K(\omega)]\right)^{2} \right\rbrace } \label{velocitycorrelation}
 \end{eqnarray}
 \end{widetext}
 Since the expression for the $y$ component position autocorrelation function is the same as that for the $x$ component, we get the same expressions for all the physical observables corresponding to the $y$ component as well. 
 Note that the above expressions for the response function, position velocity correlation function and velocity autocorrelation function are valid for any memory kernel $K(t)$. In the next section we use specific models for the memory kernel and analyse the forms of the response function, position velocity correlation function and velocity autocorrelation function for each case. The models we analyse are \cite{ford2001exact} 
 \begin{enumerate}
     \item The Ohmic model : $K(t)=2\gamma\delta(t)$
     \item The Drude model : $K(t)=\frac{\gamma}{\tau}e^{-\frac{t}{\tau}}$
 \end{enumerate}
 The Ohmic model corresponds to a memory less or Markovian bath while
the Drude model corresponds to an exponentially decaying memory ($\tau$ is the memory time) and is non-Markovian.  
 
 \section{The response function}
\subsection{Ohmic model}
In this case, the memory kernel is given by $K(\omega)=\gamma$, hence the Imaginary part of the response function (Eq. (\ref{responsefunctionomega})) is given by
 \begin{align}
\mathrm{Im}R_{x}(\omega)=\frac{\gamma}{m}\frac{(\omega^2+\omega_c^2+\gamma^2)}{\omega\left[(\omega^2+\omega_c^2+\gamma^2)^2-4\omega^2 \omega_c^2\right]}\label{imrohmic}
\end{align}
The real part $\mathrm{Re}R_{x}(\omega)$ can be obtained using the Kramers Kronig relation (Eq. (\ref{kramers})),
\begin{equation}
\begin{split}
\mathrm{Re}R_{x}(\omega)=\frac{\gamma}{m \pi}
P \int_{-\infty}^{\infty}\frac{d\omega'\big[(\omega'^2+\omega_c^2+\gamma^2)/(\omega'^2-\omega^2)\big]}{[(\omega'^2+\omega_c^2+\gamma^2)^2-4\omega'^2 \omega_c^2]} 
\end{split}
\end{equation}
Using Cauchy's residue theorem the above integral can be solved and is given by,
\begin{align}
\mathrm{Re}R_{x}(\omega)=-\frac{1}{m}\left[\frac{-\omega_c^2+\omega^2+\gamma^2}{(\omega^2+\omega_c^2+\gamma^2)^2-4\omega^2 \omega_c^2}\right]\label{reohmic}
\end{align}
The Fourier transforms of Eqs. (\ref{imrohmic}) and (\ref{reohmic}) give,
\begin{eqnarray}
\mathcal{F}(\mathrm{Im}R_{x}(\omega))&=&\frac{-i e^{-\gamma t}}{2m}\left[\frac{\omega_c \sin(\omega_c t)-\gamma \cos(\omega_c t)}{\omega_c^2+\gamma^2}\right]\nonumber\\
&-&\frac{i}{m}\left[\frac{\gamma}{\omega_c^2+\gamma^2}\right]\label{Imresponseohmic}\\
\mathcal{F}(\mathrm{Re}R_{x}(\omega))&=&\frac{e^{-\gamma t}}{2m}\left[\frac{\omega_c \sin(\omega_c t)-\gamma \cos(\omega_c t)}{\omega_c^2+\gamma^2}\right]\label{Reresponseohmic}
\end{eqnarray}
Notice that the memory kernel satisfies causality, $K(t)=0, t<0$, which implies that the integrands appearing in the Fourier transforms are analytic in the upper half plane and can have poles only in the lower half plane.

The poles lying in the lower half plane for the Ohmic case are: $\omega=\pm\omega_{c}-i\gamma$. Eqs. (\ref{Imresponseohmic}) and (\ref{Reresponseohmic}) are given by $(-2\pi i)$ times the sum of the residues at the respective poles.
Combining the above using Eq. (\ref{responset}), we get
\begin{eqnarray}
R_{x}(t)&=&\frac{1}{m}\left[\frac{\gamma-e^{-\gamma t}\left(\gamma cos(\omega_c t)-\omega_c sin(\omega_c t)\right)}{(\omega_c^2+\gamma^2)}\right]\label{responseohmic}
\end{eqnarray}
In the absence of the magnetic field, for $\omega_c\rightarrow 0$, we get the expected 
response function of a particle coupled to an Ohmic bath \cite{PhysRevE.102.062130}, which is
\begin{align}
R_{x}(t)=\frac{1}{m}\left[\frac{1-e^{-\gamma t}}{\gamma}\right] \label{responseohmic1}
\end{align}
\subsection{Drude model}
In this case, the memory kernel in the frequency domain is given by,
\begin{eqnarray}
K(\omega)&=&\frac{\gamma}{1+\omega^2 \tau^2}+i \frac{\omega \gamma \tau}{1+\omega^2 \tau^2}
\end{eqnarray}
Here $\tau$ is the memory time. Note that for $\tau\rightarrow 0$, the Drude model kernel gives the same expression as the Ohmic model kernel.
\begin{eqnarray}
\mathrm{Re}K(\omega)=\frac{\gamma}{1+\omega^2 \tau^2}, \mathrm{Im}K(\omega)=\frac{\omega \gamma \tau}{1+\omega^2 \tau^2}\label{Drudekernel}
\end{eqnarray}

Following steps similar to the ones used in the case 
of the Ohmic model, we substitute $\mathrm{Re}K(\omega)$ and $\mathrm{Im}K(\omega)$ in Eq. (\ref{responsefunctionomega}) and obtain the response function for the Drude model. The relevant integrals pertaining to the response function have been solved numerically since the expressions involved are cumbersome and thus cannot be solved analytically.

We now explore the roles of various frequencies: the cyclotron frequency $\omega_c$ and the damping rate $\gamma$. In this context we studied three different regimes:\\ 
i) Underdamped ($\omega_c >>\gamma$) ii) Overdamped\\ ($\omega_c <<\gamma$) and iii) Critically damped ($\omega_c \sim\gamma$).
\begin{figure}[htb!]
\centering
\includegraphics[width=.5\textwidth]{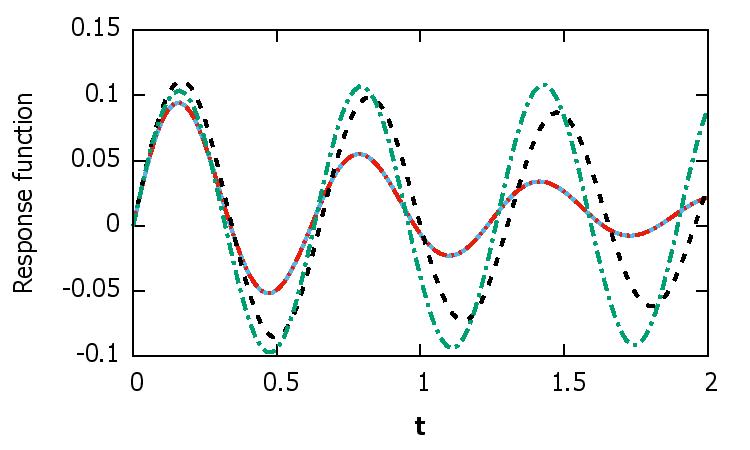} 
\includegraphics[width=.5\textwidth]{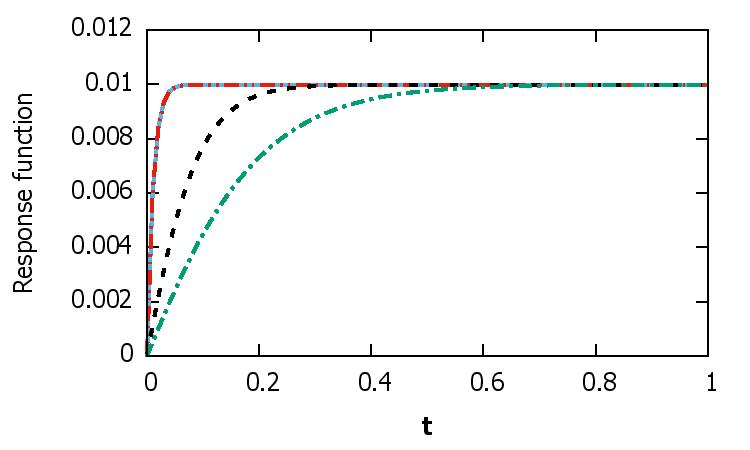} 
\includegraphics[width=.5\textwidth]{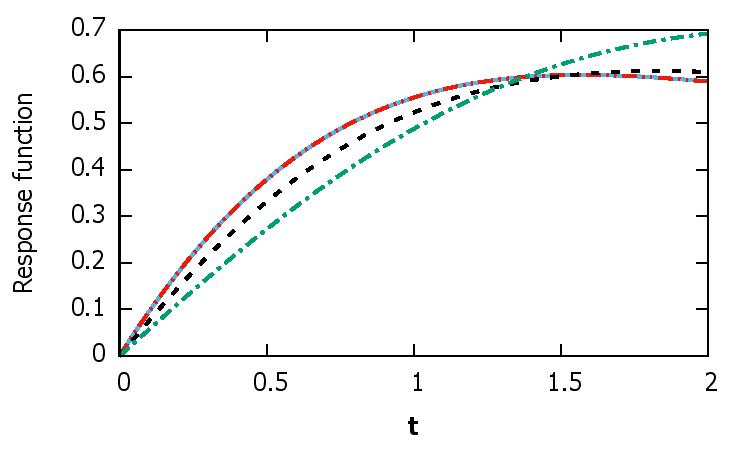} 
\caption{Time evolution of the response function,  for \textbf{Top}: the underdamped ($\omega_c$=10, $\gamma$=1). \textbf{Middle}: the overdamped ($\omega_c$=1, $\gamma$=100) \textbf{Bottom}: the critically damped ($\omega_c$=1, $\gamma$=1) cases. For all the cases the solid blue line is for the Ohmic case using Eq. (\ref{responseohmic}), the double dot-dash red line for the Drude case $\tau=0$, the dashed black line for $\tau=0.5$ and the dot-dash green line is for $\tau=1$.}\label{fig1}
\end{figure}
Fig. \ref{fig1} shows the time evolution of the response function for both the Ohmic and Drude models. For the Drude model, there is an additional timescale, the memory time $\tau$ which comes into play. We have checked the time evolution for various $\tau$ values and noticed that the presence of $\tau$
affects the quantitative features of the response function. 
\section{Position velocity correlation function}
\subsection{Ohmic model}

The position velocity correlation for the Ohmic model is given by:
\begin{align}
C_{xv_x}(t)=\frac{-i\gamma \hslash}{2 \pi m}\int_{-\infty}^{\infty}d\omega\frac{ \left(\omega^2+\omega_c^2+\gamma^2\right)\mathrm{coth}\left(\frac{\omega}{\Omega_{th}}\right)e^{-i \omega t}}{\left[\left(\omega^2+\omega_c^2+\gamma^2\right)^2-4\omega^2 \omega_c^2\right]} 
\end{align}
where $\Omega_{th}=\frac{2 k_B T}{\hslash}$ is the thermal frequency.
Using Cauchy's Residue  theorem, and choosing the lower contour consistent with causality, one can solve the above integral. The poles lying in the lower contour are $\omega=\pm\omega_{c}-i\gamma$ and $\omega=-i n\pi\Omega_{th}$, where $n=1,2,...\infty$. Here the additional poles at $\omega=-i n\pi\Omega_{th}$ pertain to the $\coth$ term. Summing over the residues, the integration yields the result\\ \\ \\

\begin{equation}
\begin{split}
C_{xv_{x}}(t)=&-\frac{\hslash e^{-\pi t \Omega_{th}}}{4 \pi m}\big[\Phi \left( e^{- \pi t \Omega_{th}},1,1+\frac{-i\omega_c +\gamma}{\pi \Omega_{th}}\right)\\
&+\Phi \left( e^{- \pi t \Omega_{th}},1,1+\frac{i\omega_c +\gamma}{\pi \Omega_{th}}\right)\\
&-\Phi \left( e^{- \pi t \Omega_{th}},1,1+\frac{-i\omega_c -\gamma}{\pi \Omega_{th}}\right)\\
&-\Phi \left( e^{- \pi t \Omega_{th}},1,1+\frac{i\omega_c -\gamma}{\pi \Omega_{th}}\right)\big]\\
&-\frac{i\hbar}{4m}e^{-\gamma t}\big[-coth\left(\frac{\omega_c+i\gamma}{\Omega_{th}}\right)e^{i\omega_c t}\\
&+coth\left(\frac{\omega_c-i\gamma}{\Omega_{th}}\right)e^{-i\omega_c t} \big] \label{PosvelDrudemain}
\end{split}
\end{equation}

where $\Phi$ is the Hurwitz-Lerch Transcendent Function, defined as:
\begin{align}
\Phi(z,s,\alpha)=\sum_{j=0}^{\infty}\frac{z^j}{(j+\alpha)^s}
\end{align}
In contrast to the response function, the position velocity correlation function has a temperature dependence due to the presence of the thermal frequency $\Omega_{th}$. In particular, we can consider the asymptotic limit of the high temperature classical domain.

In the high temperature classical domain dominated by 
thermal fluctuations, $\frac{\Omega_{th}}{\omega}>>1$. In this limit,
\begin{align}
 \Phi \left( e^{- \pi t \Omega_{th}},1,\frac{\pm i\omega_c \pm \gamma+\pi}{\pi \Omega_{th}}\right)\rightarrow \Phi \left( 0,1,\alpha\right)\rightarrow 0\\
 \mathrm{coth}\left(\frac{\omega_c \pm i\gamma}{\Omega_{th}}\right)\rightarrow \frac{\Omega_{th}}{\omega_c \pm i\gamma}
\end{align}
Hence, the position-velocity correlation turns out to be:
\begin{align}
C_{xv_x}(t)=-\frac{k_B T}{m} e^{-\gamma t}\left[\frac{\omega_c \sin(\omega_c t)-\gamma \cos(\omega_c t)}{\omega_c^2+\gamma^2} \right] 
\label{posvelhighTohmic}
\end{align}
Setting the limits $t=0$ and $\omega_c=0$, we get the following expected result: 
$C_{xv_x}(0)=\frac{k_B T}{m \gamma}$.
\\ \\


\subsection{Drude model}
\begin{figure}
\centering
\includegraphics[width=.52\textwidth]{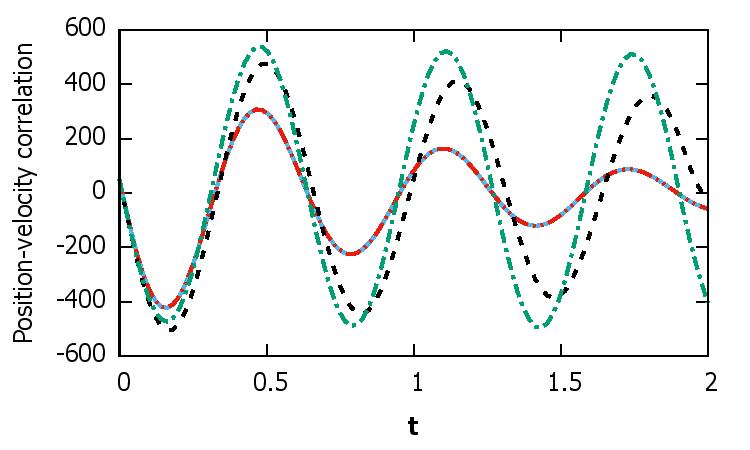} 
\includegraphics[width=.52\textwidth]{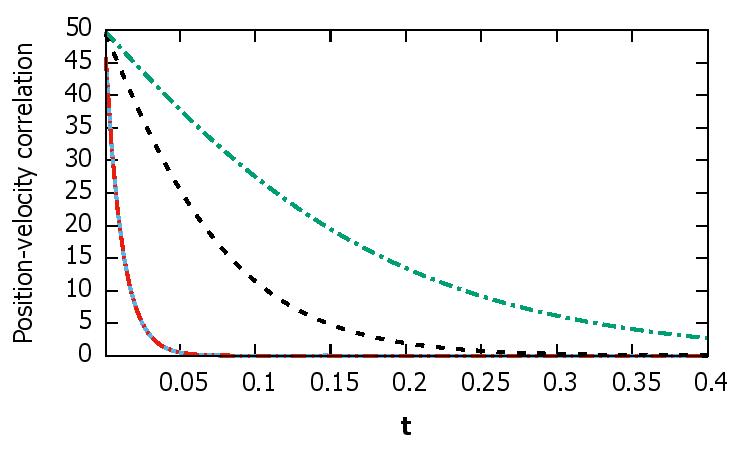}
\includegraphics[width=.52\textwidth]{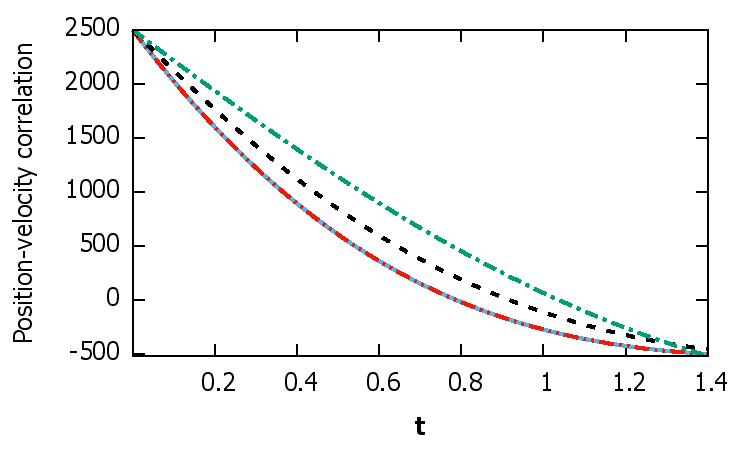}
\caption{Time evolution of the position velocity correlation function in the high temperature domain, for $\Omega_{th}=10^{4}$ \textbf{Top}: the underdamped ($\omega_c$=10, $\gamma$=1). \textbf{Middle}: the overdamped ($\omega_c$=1, $\gamma$=100) \textbf{Bottom}: the critically damped ($\omega_c$=1, $\gamma$=1)cases. For all the cases the solid blue line is for the Ohmic case using Eq. (\ref{PosvelDrudemain}), the double dot-dash red line for the Drude case $\tau=0$, the dashed black line for $\tau=0.5$ and the dot-dash green line is for $\tau=1$.}\label{fig2}
\end{figure}

\begin{figure}
\centering
\includegraphics[width=.52\textwidth]{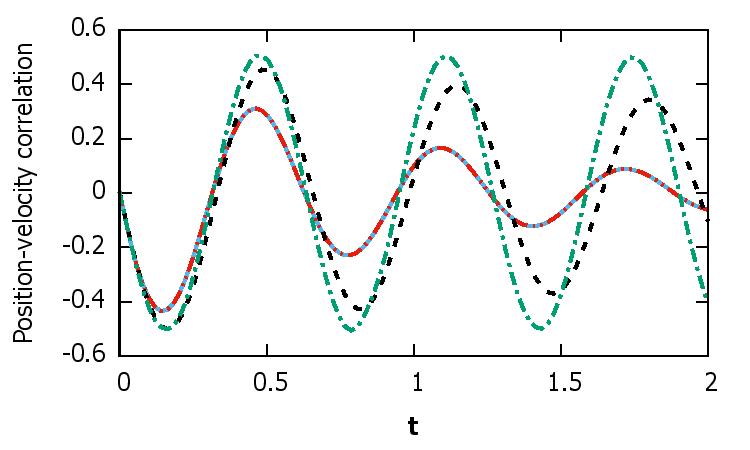} 
\includegraphics[width=.52\textwidth]{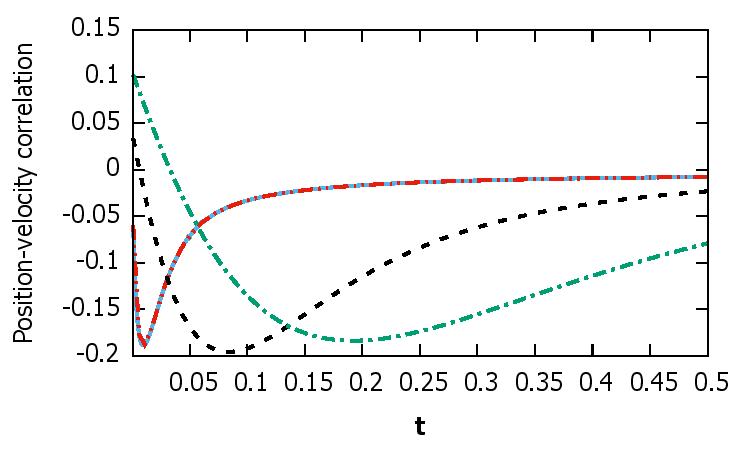} 
\includegraphics[width=.52\textwidth]{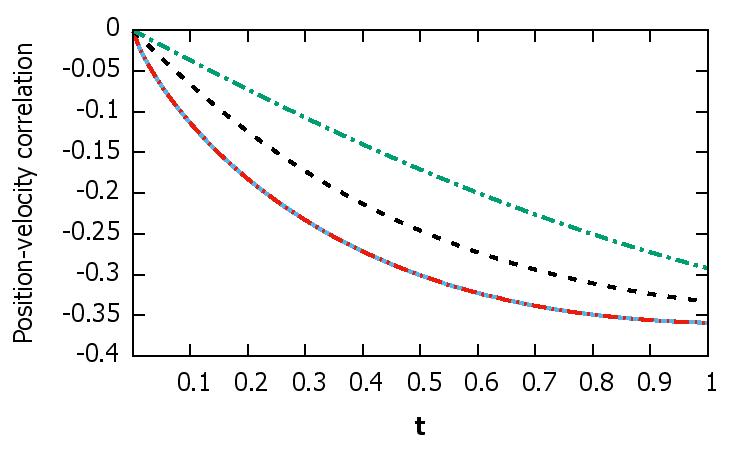} 
\caption{Time evolution of the position velocity correlation function in the low temperature domain, for $\Omega_{th}=0.1$ \textbf{Top}: the underdamped ($\omega_c$=10, $\gamma$=1). \textbf{Middle}: the overdamped ($\omega_c$=1, $\gamma$=100) \textbf{Bottom}: the critically damped ($\omega_c$=1, $\gamma$=1) cases. For all the cases solid blue line is for the Ohmic case using Eq. (\ref{PosvelDrudemain}), the double dot-dash red line for the Drude case $\tau=0$, the dashed black line for $\tau=0.5$ and the dot-dash green line is for $\tau=1$.}\label{fig3}
\end{figure}

As in the case of the Ohmic model, in this case also one can write the expression for the position velocity correlation function using Eqs. (\ref{positionvelocitycorrelation}) and (\ref{Drudekernel}). 
\begin{widetext}
\begin{equation}
\begin{split}
C_{xv_{x}}(t)=\frac{-i\hbar}{2\pi m}\int_{-\infty}^{\infty}d\omega \left(\frac{\gamma}{1+\omega^2 \tau^2}\right)\frac{\left[\left(\omega+\frac{\omega \gamma \tau}{1+ \omega^2 \tau^2}\right)^2+\omega_c^2+\left(\frac{\gamma}{1+\omega^2 \tau^2}\right)^2 \right]\coth\left(\frac{\hbar \omega}{2k_B T}\right)e^{-i \omega t}}{\left\lbrace \left[\left(\omega+\frac{\omega \gamma \tau}{1+ \omega^2 \tau^2}\right)^2+\omega_c^2+\left(\frac{\gamma}{1+\omega^2 \tau^2}\right)^2 \right]^2-4\omega_c^2\left(\omega+\frac{\omega \gamma \tau}{1+\omega^2 \tau^2}\right)^2\right\rbrace}
 \end{split}
\end{equation}
\end{widetext}
In this case too we solve the integral numerically for both the high temperature classical and low temperature quantum regimes. Figs. (\ref{fig2}) and (\ref{fig3}) show the time evolution of the position velocity correlation function in the high temperature domain and low temperature domain respectively. For both cases we have discussed the different  regimes (underdamped, overdamped and critically damped) following the same parameters and notations used in Fig. (\ref{fig1}).

 \section{Velocity autocorrelation function}
 \subsection{Ohmic model}
 For the Ohmic model, the velocity autocorrelation function is given by:
 \begin{equation}
C_{v_x}(t)=\frac{\gamma \hslash}{2 \pi m}\int_{-\infty}^{\infty}\frac{\omega(\omega^2+\omega_c^2+\gamma^2)\coth(\frac{\omega}{\Omega_{th}})}{\left[(\omega^2+\omega_c^2+\gamma^2)^2-4\omega^2 \omega_c^2\right]}e^{-i \omega t} d\omega
\end{equation}
The poles for the above expression are located at $\omega$=-in$\pi\Omega_{th}$ and $\omega=(\omega_c+i\gamma),(-\omega_c+i\gamma),(\omega_c-i\gamma)$ and $(-\omega_c-i\gamma)$. Out of these, only the poles at -in$\pi\Omega_{th}$, $(\omega_c-i\gamma)$ and $(-\omega_c-i\gamma)$ lie within the lower contour. So, the above integration can be calculated using Cauchy's residue theorem and we get:
\begin{equation}
\begin{split}
C_{v_x}(t)=&-\frac{i\hslash e^{-\pi t \Omega_{th}}}{4 \pi m}\times\\
&\bigg\lbrace (\omega_c+i\gamma) \bigg[\Phi \left( e^{- \pi t \Omega_{th}},1,1+\frac{-i\omega_c +\gamma}{\pi \Omega_{th}}\right)\\
&+\Phi \left( e^{- \pi t \Omega_{th}},1,1+\frac{i\omega_c -\gamma}{\pi \Omega_{th}}\right)\bigg]\\
&-(\omega_c-i\gamma)\bigg[\Phi \left( e^{- \pi t \Omega_{th}},1,1+\frac{i\omega_c +\gamma}{\pi \Omega_{th}}\right)\\
&-\Phi \left( e^{- \pi t \Omega_{th}},1,1+\frac{-i\omega_c -\gamma}{\pi \Omega_{th}}\right)\bigg]\bigg\rbrace\\
&+\frac{\hbar}{4m}e^{-\gamma t}\bigg[(\omega_c+i \gamma)coth\left(\frac{\omega_c+i\gamma}{\Omega_{th}}\right)e^{i\omega_c t}\\
&+(\omega_c-i \gamma)coth\left(\frac{\omega_c-i\gamma}{\Omega_{th}}\right)e^{-i\omega_c t} \bigg]
\end{split}\label{velautocorrDrudemain}
\end{equation}

In the high temperature limit ($\frac{\Omega_{th}}{ \omega}>>1$), 
the velocity auto-correlation turns out to be
\begin{align}
C_{v_x}(t)=\frac{k_B T}{m}e^{-\gamma t} \cos(\omega_c t)
\label{velhighTohmic}
\end{align}
At $t=0$,  $ C_{v_x}(0)=\frac{k_B T}{m}$, which is expected in the thermal fluctuation dominated classical regime and is consistent with the Equipartition Theorem. 

In Ref. \cite{sushantapramana} the authors evaluate 
the velocity autocorrelation for a charged particle in
a magnetic field for an Ohmic model. We go beyond that 
in studying in detail the interplay of various different time scales as revealed in the plots and related discussion
in subsequent sections. 

\subsection{Drude model}
Using Eqs. (\ref{velocitycorrelation}) and (\ref{Drudekernel}), the velocity autocorrelation in this case is given by,
\begin{widetext}
\begin{equation}
\begin{split}
C_{v_{x}}(t)=\frac{\hbar}{2\pi m}\int_{-\infty}^{\infty}d\omega \left(\frac{\gamma \omega}{1+\omega^2 \tau^2}\right)\frac{\left[\left(\omega+\frac{\omega \gamma \tau}{1+ \omega^2 \tau^2}\right)^2+\omega_c^2+\left(\frac{\gamma}{1+\omega^2 \tau^2}\right)^2 \right]\coth\left(\frac{\hbar \omega}{2k_B T}\right)e^{-i \omega t}}{\left\lbrace \left[\left(\omega+\frac{\omega \gamma \tau}{1+ \omega^2 \tau^2}\right)^2+\omega_c^2+\left(\frac{\gamma}{1+\omega^2 \tau^2}\right)^2 \right]^2-4\omega_c^2\left(\omega+\frac{\omega \gamma \tau}{1+\omega^2 \tau^2}\right)^2\right\rbrace}
 \end{split}
\end{equation}
\end{widetext}
We analyse this integral numerically like the other physical observables discussed in earlier sections, for the Drude model, for both the high temperature and low temperature regimes. 
 
 \begin{figure}
\centering
\includegraphics[width=.52\textwidth]{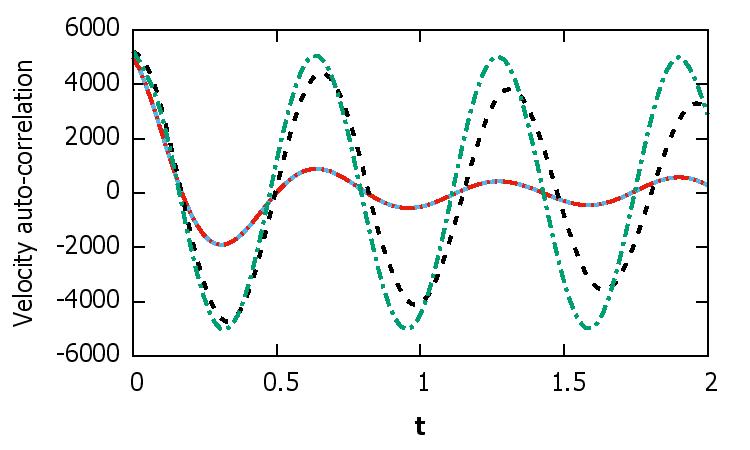} 
\includegraphics[width=.52\textwidth]{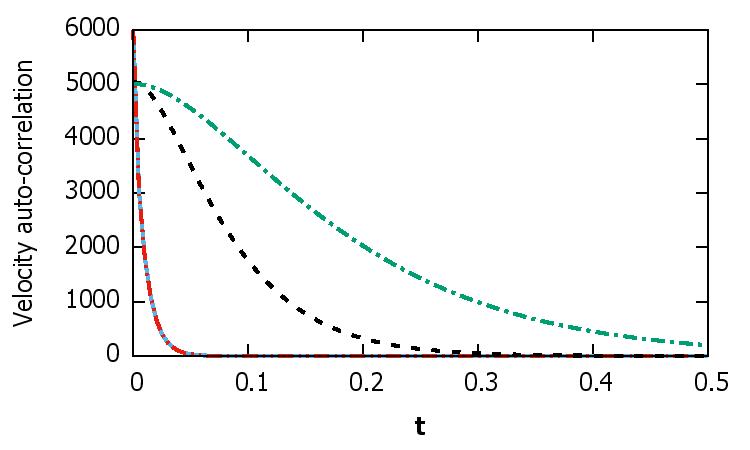} 
\includegraphics[width=.52\textwidth]{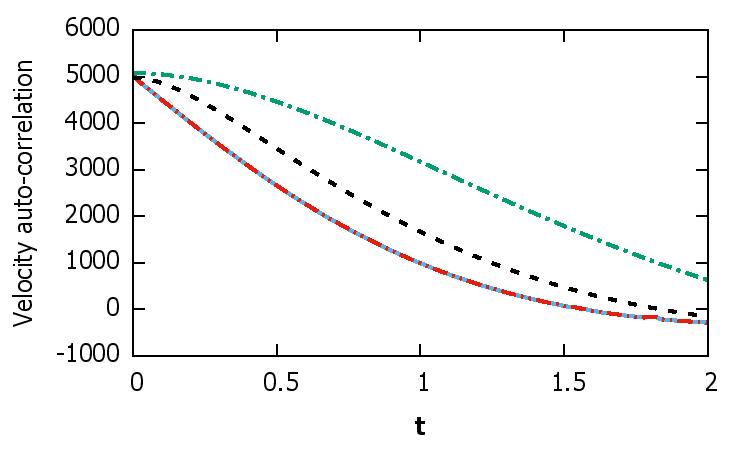} 
\caption{Time evolution of the velocity auto-correlation function in the high temperature domain, for $\Omega_{th}=10^{4}$ \textbf{Top}: the underdamped ($\omega_c$=10, $\gamma$=1). \textbf{Middle}: the overdamped ($\omega_c$=1, $\gamma$=100) \textbf{Bottom}: the critically damped ($\omega_c$=1, $\gamma$=1) cases. For all the cases solid blue line is for the Ohmic case using Eq. (\ref{velautocorrDrudemain}), the double dot-dash red line for the Drude case $\tau=0$, the dashed black line for $\tau=0.5$ and the dot-dash green line is for $\tau=1$.}\label{fig4}
\end{figure}

 \begin{figure}
\centering
\includegraphics[width=.52\textwidth]{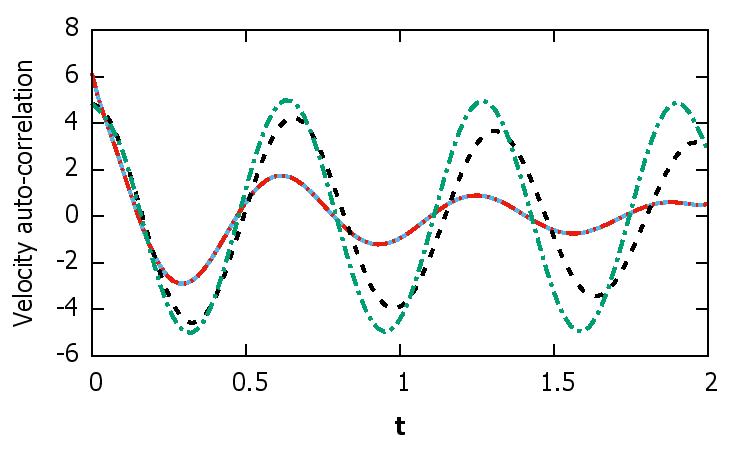} 
\includegraphics[width=.52\textwidth]{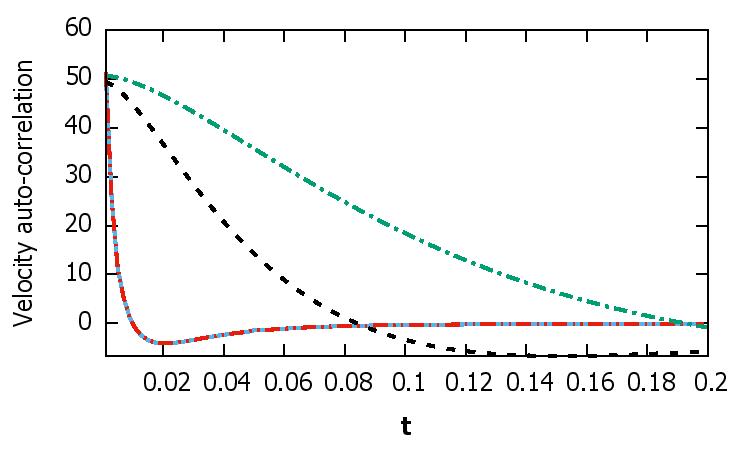} 
\includegraphics[width=.52\textwidth]{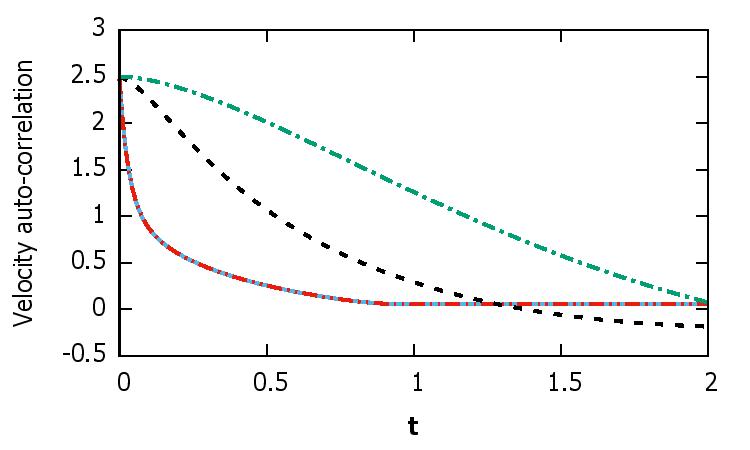} 
\caption{Time evolution of velocity auto-correlation function in the low temperature domain, for $\Omega_{th}=0.1$ \textbf{Top}: the underdamped ($\omega_c=10$, $\gamma=1$). \textbf{Middle}: the overdamped ($\omega_c=1$, $\gamma=100$) \textbf{Bottom}: the critically damped ($\omega_c=1$, $\gamma=1$). For all the cases the solid blue line is for the Ohmic case using Eq. (\ref{velautocorrDrudemain}), double dot-dash red line for the Drude case $\tau=0$, the dashed black line for $\tau=0.5$ and the dot-dash green line is for $\tau=1$.}\label{fig5}
\end{figure}

\vspace*{0.2cm}
Figs. (\ref{fig4}) and (\ref{fig5}) show the time evolution of the velocity autocorrelation function in the high temperature domain and low temperature domain respectively. For both cases we have discussed the different  regimes (underdamped, overdamped and critically damped) following the same parameters and notations used in Figs. (\ref{fig2}) and (\ref{fig3}).
\section{Results: Comparison of Ohmic and Drude models}

In this section we discuss the results displayed in Figs. (\ref{fig1}) - (\ref{fig5})

In Fig. (\ref{fig1}) we study the 
behaviour of the response function for the Ohmic model and for the Drude model for various values of $\tau$, the memory time
(the Drude time) in the 
underdamped ($\omega_c >> \gamma$), overdamped ($\omega_c << \gamma$) and 
critically damped ($\omega_c = \gamma$) regimes. We have explicit analytical forms for the Ohmic model (Eq. (\ref{responseohmic}), Eq. (\ref{PosvelDrudemain}) 
and Eq. (\ref{velautocorrDrudemain}) 
)
which enables an easy comparison with the results obtained in the Ohmic limit ($\tau=0$) for the Drude model. We notice 
that as $\tau$ goes up, the oscillations in the underdamped regime get to be more sustained. In the overdamped regime, for the Ohmic case there is an initial steep rise in the response function before settling down to the constant value of $\frac{1}{m \gamma}$ (See Eq. (\ref{responseohmic1})). This behaviour
of the response function $R(t)$ of an 
initial increase followed by saturation to a fixed  value determined by the viscous damping rate can be understood as follows. When we perturb a Brownian
particle, it initially has a directional displacement
before it gets completely randomised by a rate 
determined by the viscous
damping rate. The overall trend is the same for the Drude model too. However, we notice that the Drude time scale $\tau$ leads to a more sustained oscillation in the underdamped regime and leads to an effective slowing down of the damping in the overdamped regime. This can be understood as follows. The presence of memory, characterized by $\tau$ in the Drude model leads to a slowing down of the rise of the response function (as revealed in Fig.\ref{fig1}), which in turn means that randomising of the motion of the Brownian particle is slowed down leading to a weaker effect of dissipation on the particle. Since the damped harmonic oscillation is a consequence of a competition between the oscillatory effect of the cyclotron frequency and the damping effect of viscosity, it is clear that the presence of a finite $\tau$ results in more sustained oscillations compared to the Ohmic case where $\tau=0$.   In the critically damped case , the behaviour of the response function is in between the two extremes of underdamped and overdamped regimes. 

Let us now discuss the position-velocity correlation function. 
We display a family of curves for various values of $\tau$ for the position-velocity correlation function in the high temperature regime for the overdamped, underdamped and critically damped cases in Fig. \ref{fig2}. We notice a  transition from an oscillatory behaviour to an overdamped monotonic behaviour stemming from a competition between the oscillatory time scale set by the cyclotron frequency $\omega_c$ and the damping rate $\gamma$. Such a transition from an oscillatory to a damped monotonic behaviour has been noticed in some earlier works in the context of damping of Bloch oscillations in optical lattices \cite{Kolovsky}. However, in Ref.\cite{Kolovsky} the origin of these damped harmonic oscillations is quite distinct from the context dealt with here in this paper. Ref.\cite{Kolovsky} studies Bloch oscillations of cold neutral atoms in an optical lattice. Spontaneous emission causes the decay of Bloch oscillations, the  decay rate being set by the rate of spontaneous emission. Furthermore, from our results, we notice that similar to the case of response function, in the under-damped regime, the increase in the memory time scale leads to more sustained oscillations. In addition, in the over-damped and critically-damped regime, the damping is slower for larger memory scales as seen in Fig. \ref{fig2}.

As in the high temperature domain, we find that even in the low temperature domain the position-velocity correlation function exhibits oscillations in the under-damped regime (Fig. \ref{fig3}). What is somewhat surprising is the presence of a non-monotonic trend in the position-velocity function in the over-damped regime at low temperatures. This is true for both the Drude case and the Ohmic case ($\tau=0$) limit. This can be explained by the fact that noise correlations are inherently non-trivial (non-Markovian) in the low temperature quantum domain. This is, perhaps, the origin of sustained memory induced non-monotonic trends in the curves in the low temperature over-damped regime. The position-velocity correlation goes towards negative values in the beginning and then turns around and finally tends to zero at large times. The time at which this turn around takes place is dependent on the memory time $\tau$ and 
shifts towards larger values of time as $\tau$ increases. It is interesting  to note that there has been a similar observation of sustained oscillations in the context of a classical Langevin equation (valid in the high temperature domain) of a charged particle in a magnetic field for a Drude model in the presence of an exponentially correlated noise \cite{Paraan}. Nontrivial noise correlations is a common feature in these two disparate cases (a Classical Langevin dynamics of a charged particle in a magnetic field with correlated noise \cite{Paraan} and a Quantum Langevin dynamics of a charged particle in a magnetic field with non-trivial quantum correlated noise, dealt with in this paper). Notice that in our analysis in the high temperature limit of $\frac{\beta\hbar \omega}{2} << 1$, such non-trivial noise correlations are absent, simply because one has a memory free delta correlated noise in that limit and thus we obtain a monotonic behaviour starting from a large positive value of the position-velocity correlation function which gets damped and goes over to zero in the long time limit.

The trends followed by the velocity autocorrelation 
function in the various time-temperature regimes (Fig. \ref{fig4} and Fig. \ref{fig5}) are similar to that of the position-velocity 
autocorrelation function. However, the non-monotonic
feature discussed in the low temperature, overdamped regime 
is less pronounced in the case of the velocity autocorrelation
function compared to the case of the position-velocity correlation function. 

 \section{Experimental implications}
In recent experiments correlation functions have been studied using a variety of methods. Many of these experiments measure the properties of the anomalous diffusion of cold atoms in optical lattices \cite{Barkai,Katori,Sagi}. The experimental technique involved in ref. \cite{Grimm} uses optical tweezers for trapping the particles and the position correlations of the particles are measured using a high-bandwidth photodetector.  In ref. \cite{posvelprl}, the  position-velocity correlation function has been studied for ultracold Rb atomic cloud undergoing anomalous diffusion. The position velocity correlation is then measured using a tomographic method which is a combination of absorption imaging and Raman velocity selection. In a recent paper \cite{bhar2021}, the measurement of the response function of ultracold Rb atomic cloud in a magneto optical trap (MOT) and the spatial diffusion of the cloud in the absence of the MOT have been presented and analysed. As we know from the linear response theory, the response function is a measure of how a system responds to an external drive. In the experiment \cite{bhar2021} a pulsed homogeneous magnetic field is used as an external driving force for measuring the response function. The atomic cloud gets displaced due to the external drive. Then it is allowed to equilibriate for some time and then it finally returns to its original position when the external drive is turned off. The position of the cold atomic cloud is recorded at regular intervals of time after turning off the external drive and thus the path of the cloud is traced. In the same experiment, spatial diffusion of the atomic cloud is measured in the viscous medium provided by optical molasses. In this case the MOT is turned off and the atomic cloud is allowed to diffuse in the presence of optical molasses and the mean square displacement (the position correlation function) is measured using an absorption imaging technique. As discussed in section II, a knowledge of the position autocorrelation enables us to compute the position velocity and velocity autocorrelation functions. These predictions can be tested using the techniques of ref. \cite{bhar2021} with hybrid traps for ions and neutral atoms, and additionally a uniform magnetic field can be provided using a combination of Helmholtz configuration magnetic coils. 
\section{Concluding Remarks}
 In this paper we study the interplay between 
 the cyclotron frequency and the viscous damping rate
 via the Quantum Langevin Equation for a 
 charged particle coupled to a bath in the presence
 of a magnetic field. In particular, we
 study the response function, the position-velocity correlation function and the velocity autocorrelation for two bath models- the Ohmic bath and the Drude bath and make a detailed comparison in various time-temperature regimes. Here it is pertinent to mention that, such a detailed investigation has not been done earlier to our knowledge. Specifically, our results of position-velocity correlation are of extreme importance, as it has been rarely studied analytically in literature, in spite of having experimental relevance \cite{posvelprl}.

In the zero point fluctuation dominated low temperature regime, non-trivial noise correlations lead to some interesting features in the transition from an oscillatory
to a monotonic behaviour.
We also study the role of the memory time scale which comes into play in the Drude model and investigate the effect
of this additional time scale. We thus see a rich interplay of various different time scales set by the cyclotron frequency $\omega_c$, the damping rate $\gamma$, the Drude time scale $\tau$ and the thermal time scale $\beta \hbar$ which controls the noise correlations in the quantum domain. Our study is unique in addressing the richness of all these time scales, sweeping across the thermal fluctuation dominated classical regime and the quantum fluctuation dominated quantum regime. We are not aware of such a comprehensive study  of the response function, the position-velocity correlation function and the velocity auto-correlation function for a charged particle in a magnetic field in a viscous medium.

 In ref.\cite{Paul,Felderhof,Aslangul,Pottier,Jung}, a power law decay (t$^{-\alpha}$) of the velocity autocorrelation function has been observed at long times, with $\alpha$=3/2 in the classical regime and $\alpha$=2  in the quantum regime. In  our study too we notice a slow long time decaying behaviour [Fig. \ref{fig4} and Fig. \ref{fig5}] of the correlation functions at over damped and critically damped regimes, both in the classical and quantum domains. The precise nature of the long time tails lies outside the scope of the present study and can be analyzed in more detail in the future.

In this paper we have considered a quantum Lange-vin equation (QLE) of a charged quantum Brownian particle in the presence of a magnetic field and linearly coupled via position coordinate to a bath. 
In contrast, in Ref. \cite{malay} the authors derived a quantum Langevin equation for a charged quantum particle in a
harmonic potential in the presence of a uniform external
magnetic field and coupled linearly through the momentum variables to a bath of oscillators. 
In that context they noticed that the magnetic field appears through a quantum generalized 
classical Lorentz force term. In addition, the QLE involves a random force independent of the magnetic field. 
While these aspects are also there in the present analysis of the QLE for a charged
quantum Brownian particle coupled to the bath via position coordinate \cite{Li2}, there are significant differences 
listed below:
(i) The random force has a modified form with symmetric correlation and unequal time commutator different
from those in the case of coordinate-coordinate coupling,
(ii) the inertial term and the harmonic potential term in
the QLE get renormalized as a consequence of the renormalization of the mass, and 
(iii) the memory function characterizing the mean force in the QLE not only has a
magnetic field-independent diagonal part, but also an explicit magnetic field dependent off-diagonal part \cite{malay}. We therefore expect the behaviour of the various correlation functions analysed and depicted here to get 
qualitatively and quantitatively modified in the context of a momentum coupling between the charged particle and the bath.
We expect our work to generate interest amongst  experimentalists to test the predictions that 
stem out of our theoretical study.


 \bibliography{supurnareferences}
 
  \end{document}